# Hardware Acceleration of Block-Diffusion LLM for Edge Devices

Wei-Hsing Huang, Kiseok Lee, Ming-Yen Lee, Weiyu Sun, Cheng-Jhih Shih, Gayatri Tanksali, Arpit Khandelwal, Pin-Jun Chen, Yingyan (Celine) Lin and Shimeng Yu, Fellow, IEEE

***Abstract*—Single-stream (batch-one) edge inference cannot amortize weight traffic across requests. Full-attention diffusion LLMs recompute the entire sequence at every step; native block diffusion makes completed blocks immutable and exactly cacheable, yet refinement still streams prefix KV and FFN weights. We co-design WIFiV-LPDDR, a wide-I/O LPDDR system for precision-tagged reads, BRQ-KV for a canonical low-rank-plus-INT8-residual prefix with query-dependent per-entry precision, and DAT-FFN for drift-mapped canonical replacement, adjacent-stage-corrected low-bit delta, or cached-state carry while keeping live activations unquantized. Both map to an input-stationary mixed-precision systolic array. For the evaluated 1.5B/7B models on modeled Jetson-class platforms, the full stack provides arithmetic-mean energy-reduction factors of 3.79× / 3.96× and arithmetic-mean latency speedups of 2.88× / 4.44× at the reported DAT-FFN settings; every corresponding compressed model-benchmark score drops by less than one absolute percentage point from its baseline.**

***Index Terms*—Large language models, Diffusion models, Edge AI, Hardware acceleration.**

## I. INTRODUCTION

We study private batch-one LLM inference on personal devices, where cross-request batching is unavailable and autoregressive decoding exposes little intra-request weight reuse [1]. Full-attention dLLMs predict masked positions in parallel but keep the response mutable [2]; Fast-dLLM v2 makes completed blocks immutable and exactly cacheable while refining only the active block [3]. Representative dialogue and reasoning workloads generate multi-token responses [1]. Since prefill contributes at most 3% of measured latency and energy (Section 4.B), we target repeated block-diffusion refinement rather than prefill.

Refinement repeatedly streams immutable prefix KV and static FFN weights from DRAM (Figure 1). We co-design WIFiV-LPDDR (Wide-I/O FinFET VCT LPDDR), an LPDDR-generation-agnostic organization combining $4F^2$ vertical-channel-transistor (VCT)/CMOS-bonded-array (CBA) memory, FinFET peripheral logic, and fine-pitch 2.5D I/O [4][5]; we model an LPDDR5 instantiation for Jetson Orin Nano/NX [6]. Our framework tiers persistent state while keeping live Q/K/V and FFN activations in BF16: KV tiers control entry-read precision, and FFN tiers select replacement, low-bit delta, or carry.

BRQ-KV (Block-Refreshed Query-Ranked KV) keeps a canonical low-rank-plus-8-bit (q8) residual prefix, retains every entry through its low-rank base, and reversibly reassigns residual precision with block queries. DAT-FFN (Drift-Adaptive Tiered FFN) reuses a block-entry channel order while drift moves replacement/delta/carry boundaries. Both map to an input-stationary mixed-precision array. We present an end-to-end co-design of KV/FFN precision policies and a precision-tagged wide-I/O DRAM system for text block-diffusion LLM inference.

Across five workloads at the reported DAT-FFN settings, the full stack provides arithmetic-mean energy-reduction factors of 3.79× / 3.96× and arithmetic-mean latency speedups of 2.88× / 4.44× for 1.5B/7B, while every corresponding compressed benchmark score drops by less than one absolute percentage point from its baseline.

Our contributions are:

- WIFiV-LPDDR and an end-to-end memory–accelerator–algorithm stack with precision-tagged delivery for native block diffusion.
- BRQ-KV, which preserves a recoverable canonical prefix while reassigning per-entry residual precision; its lowest tier remains base-only.
- DAT-FFN, which maps drift to replacement/delta/carry, refreshes the nonzero-tier anchor, and reuses adjacent q8→q4→q2 corrections.

## II. BACKGROUND AND MOTIVATION

### A. Single-Request and Native Block Diffusion

A single foreground request exposes little AR weight reuse because each decoding step advances only one token [1]. Full-attention masked dLLMs create intra-request parallelism but keep all response positions mutable [2]; this prevents exact persistent-prefix caching. Fast-dLLM v2 instead makes blocks causal model units: completed blocks form an immutable prefix, while bidirectional refinement is confined to the active block [3]. Although block diffusion offers intra-request parallelism, repeated steps still reread weights and prefix KV. This immutable-prefix/active-block split is the temporal structure exploited by BRQ-KV and DAT-FFN.

This work was supported in part by PRISM, one of the SRC/DARPA JUMP 2.0 centers. (Corresponding author: Shimeng Yu, e-mail: shimeng.yu@ece.gatech.edu). Wei-Hsing Huang, Kiseok Lee, Ming-Yen Lee, Weiyu Sun contributed equally to this work.

Wei-Hsing Huang, Kiseok Lee, Ming-Yen Lee, Gayatri Tanksali, Pin-Jun Chen and Shimeng Yu are with the School of Electrical and Computer Engineering, Georgia Institute of Technology, Atlanta, GA 30332 USA.

Weiyu Sun, Cheng-Jhih Shih, Arpit Khandelwal and Yingyan (Celine) Lin are with the School of Computer Science, Georgia Institute of Technology, Atlanta, GA 30332 USA.

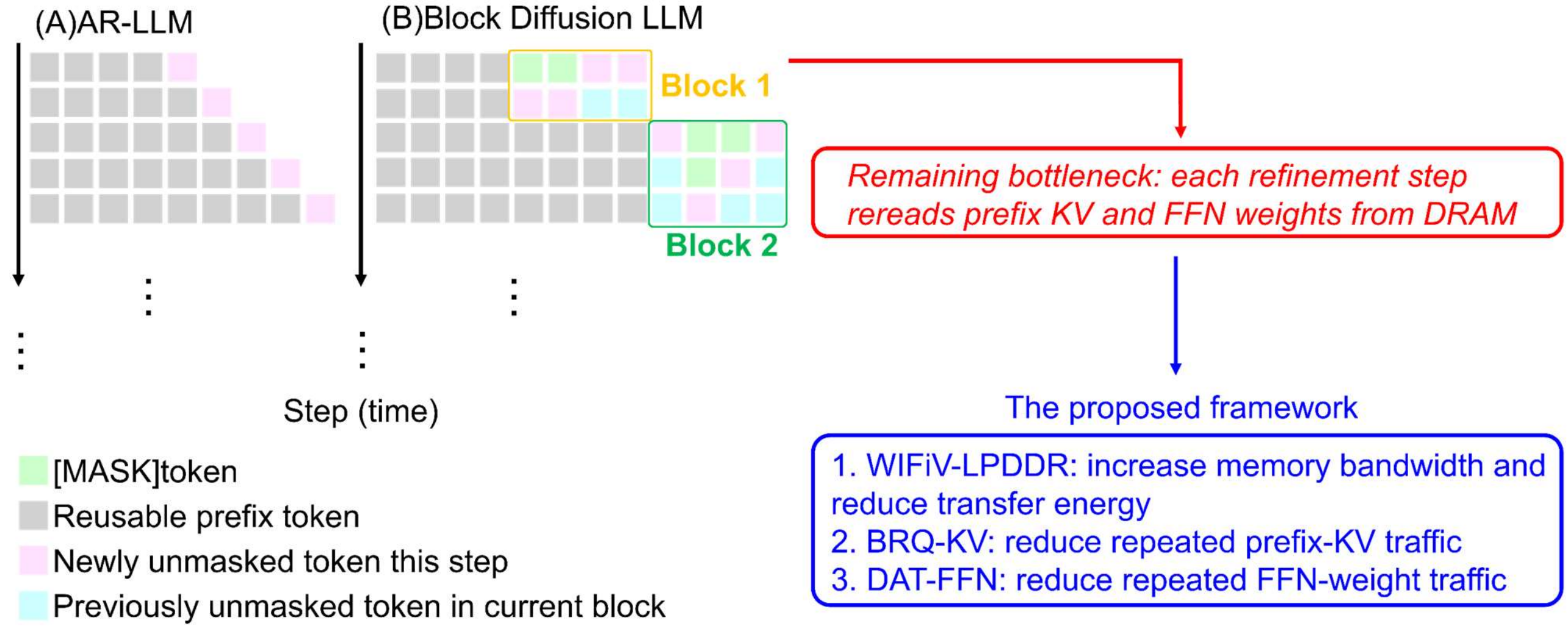

**Fig. 1.** The shift from autoregressive decoding to native block diffusion at batch size one, and the per-step prefix-KV and FFN-weight DRAM traffic that motivates the proposed framework.

### *B. Memory Systems for Future Edge AI*

$4F^2$ VCT technology extends DRAM density, while peri-under-cell, cell-on-peripheral, and CBA architectures decouple the array and peripheral processes [4]. This separation permits advanced peripheral logic: reported FinFET technologies improve drive current by roughly 30–38%, reducing the first-order delay $t_{\mathrm{pd}} \propto C_{\mathrm{eff}} V_{\mathrm{DD}} / I_{\mathrm{drive}}$. Fine-pitch 2.5D integration provides denser die-to-die wiring than conventional package-on-package (PoP) interfaces [5]. HBM-class memory marks the high-bandwidth end of the design space. For edge AI, WIFiV-LPDDR instead scales an LPDDR-family interface through package-enabled I/O width and stronger peripheral logic; Section 4.A reports the evaluated LPDDR5-based instantiation. This stronger memory hardware fabric complements rather than replaces the algorithm techniques in Sections 3.C and 3.D, which reduce the repeated KV and FFN payload requested by block diffusion.

### *C. Related KV and FFN Policies*

MAGE selects layer- and head-specific binary top-$k$ subsets of historical-prefix KV entries at each block's first All-[MASK] step and reuses them for the block's remaining steps [7]. This reduces KV reads, but participation remains binary: entries outside the subset contribute nothing to sparse-attention steps, with no intermediate participation level. MAGE also leaves FFN unchanged, so under our weight-streaming edge model each step still streams all FFN weights from DRAM and incurs their data-movement energy.

Chipmunk targets image/video Diffusion Transformers (DiTs) [8]. It periodically refreshes an active-neuron mask by projecting token-group means through the full first-layer matrix $W_1$ and applying top-$k$. Under our weight-streaming model, selection incurs a recurring full-$W_1$ read, and unselected neurons receive no current-step correction until reselection or a dense refresh. DSTAR targets mixed-precision differential activations in visual DiTs rather than persistent FFN-weight delivery [9]; under our weight-streaming model, those weights are reread each step in one fixed representation.

BRQ-KV instead assigns reversible q8/q4/q2/q0 residual precision while retaining every historical entry through its low-rank base; no entry is evicted, and lower-tier entries may later be promoted (Section 3.C).

DAT-FFN forms one channel ordering per decode block from canonical block-entry state and offline down-projection norms, then reuses it while drift moves only replacement/view-delta/carry boundaries. Recurrent-state refresh after each interior step needs neither reranking nor an additional full-matrix projection (Section 3.D).

## III. METHOD

### *A. Notation and Decoding Setting*

Native block diffusion uses outer block size $B$ and optional sub-block size $S$ [3]. A forward pass is decoding *step* $t$; phase roles are *block entry* (the first step on a new block), *interior refinement*, and *block completion* (the finalizing step). Each step updates $T \in \{S, B\}$ positions, defining a *query span* $s$: an $S$-token sub-block or the whole $B$-token block. Steps on the same span are *compatible*; only compatible steps may reuse that span's tier map (Section 3.C). Completed context is immutable, the active block remains mutable, and sub-block-free backbones set $T = B$; any block-diffusion LLM is supported.

q8 is the only persistent integer code; q4/q2 are read-time views. The tier budget $\boldsymbol{\rho} = (\rho_8, \rho_4, \rho_2, \rho_0)$ satisfies $\sum_p \rho_p = 1$ and applies to KV entries or FFN channels. KV q0 retains its low-rank base but omits the residual; FFN q0 carries cached state without a current update. Concrete formats and budgets appear in Section 4.

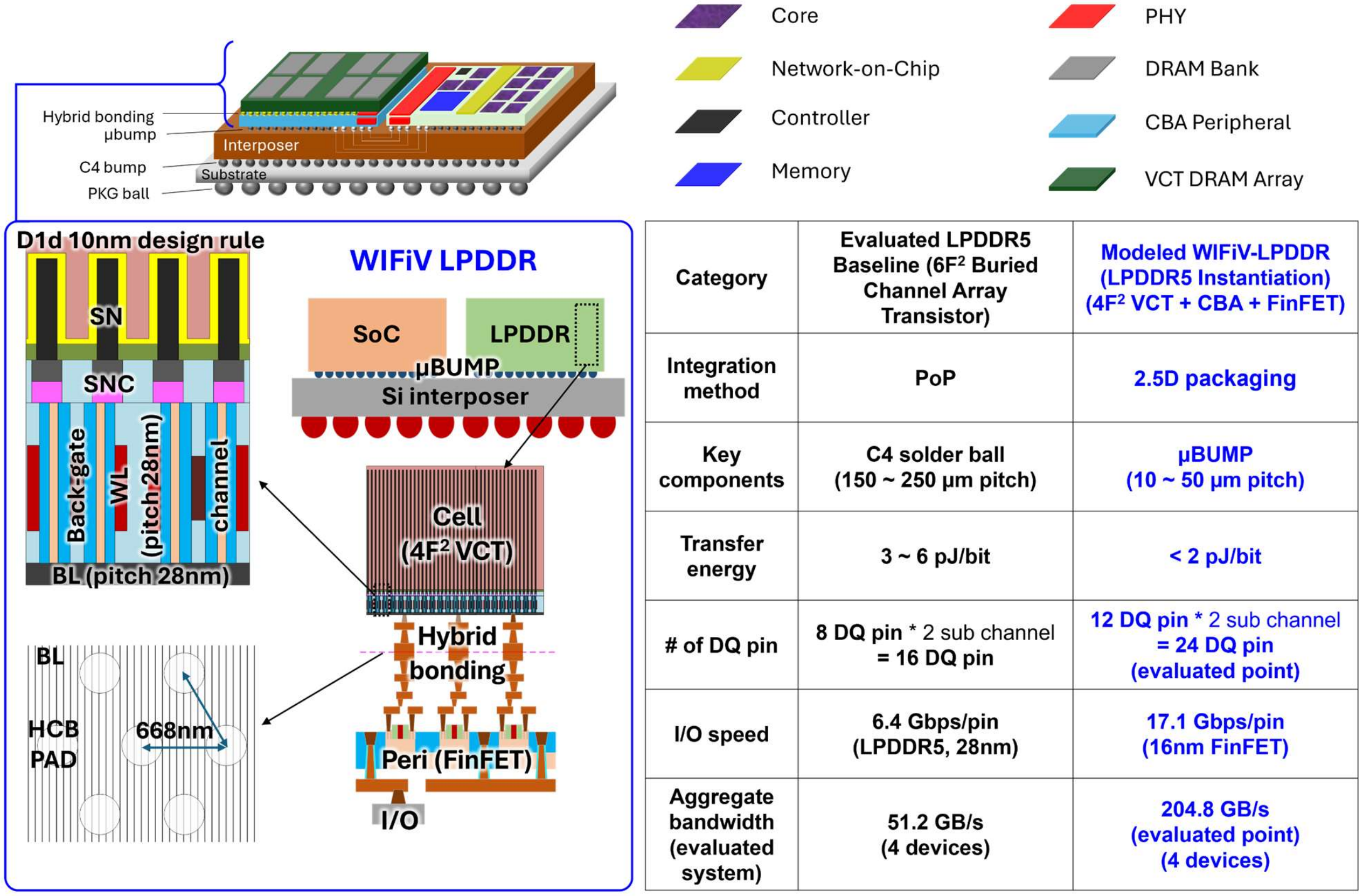

| Category | Evaluated LPDDR5 Baseline (6F² Buried Channel Array Transistor) | Modeled WIFiV-LPDDR (LPDDR5 Instantiation) (4F² VCT + CBA + FinFET) |
|---|---|---|
| Integration method | PoP | 2.5D packaging |
| Key components | C4 solder ball (150 ~ 250 μm pitch) | μBUMP (10 ~ 50 μm pitch) |
| Transfer energy | 3 ~ 6 pJ/bit | < 2 pJ/bit |
| # of DQ pin | 8 DQ pin * 2 sub channel = 16 DQ pin | 12 DQ pin * 2 sub channel = 24 DQ pin (evaluated point) |
| I/O speed | 6.4 Gbps/pin (LPDDR5, 28nm) | 17.1 Gbps/pin (16nm FinFET) |
| Aggregate bandwidth (evaluated system) | 51.2 GB/s (4 devices) | 204.8 GB/s (evaluated point) (4 devices) |

**Fig. 2.** Generation-agnostic WIFiV-LPDDR organization and an LPDDR5-based modeled instantiation. Evaluation uses a conservative 4× bandwidth point.

### *B. WIFiV-LPDDR Architecture and Mixed-Precision Data Delivery*

**Device, circuit, and package co-design.** We propose a WIFiV-LPDDR customization that combines a dense $4F^2$ VCT cell array, a CBA with FinFET peripheral circuits for faster peripheral logic, and 2.5D integration with the compute die over a silicon interposer [4][5]. The CBA boundary allows the array and peripheral logic to use independent process and thermal budgets, while the fine-pitch microbump interface converts package area into effective DQ width. As summarized in Figure 2, WIFiV-LPDDR is an LPDDR-generation-agnostic wide-I/O organization that scales aggregate bandwidth through package-enabled DQ width and stronger peripheral logic supporting a higher per-pin rate. The figure shows the LPDDR5-based instantiation evaluated in this work; Eq. (2) parameterizes other instantiations through $N_{\mathrm{DQ}}$, $R_{\mathrm{pin}}$, and $\eta_{\mathrm{link}}$. The fine-pitch 2.5D interface also reduces interconnect energy per bit, enabling this bandwidth scaling without the data-transfer energy cost of conventional package-on-package interfaces. We model peripheral propagation delay with

$$t_{\mathrm{pd}} \propto \frac{C_{\mathrm{eff}} V_{\mathrm{DD}}}{I_{\mathrm{drive}}}, \tag{1}$$

and peak payload bandwidth with

$$\mathrm{BW}_{\mathrm{peak}} = \frac{\eta_{\mathrm{link}} N_{\mathrm{DQ}} R_{\mathrm{pin}}}{8}, \tag{2}$$

where $N_{\mathrm{DQ}}$ is the package-enabled effective I/O width, $R_{\mathrm{pin}}$ is the modeled per-pin transfer rate, and $\eta_{\mathrm{link}}$ captures protocol and scheduling efficiency; $C_{\mathrm{eff}}$, $V_{\mathrm{DD}}$, and $I_{\mathrm{drive}}$ denote effective switched peripheral capacitance, supply voltage, and drive current. The cited VCT/CBA and 2.5D-interconnect results provide technology anchors for this design.

**Memory-side precision conversion.** WIFiV-LPDDR stores canonical q8 residual or weight codes and accepts a precision tag. A nonzero read moves the 8-bit word to a local buffer and emits only the requested payload. KV q4/q2 use calibrated clamp/bit selection; FFN q4 uses clamp, rounding shift, and saturation, while ternary q2 uses two signed comparisons and a mux. Runtime uses fixed clamp/shift logic, comparators, a precision decoder, and muxes; scales are applied on chip. For payload precision $p$, the first-order movement-energy model is

$$E_{\mathrm{move}}(p) = \begin{cases} 8e_{\mathrm{core}} + p e_{\mathrm{down}} + E_{\mathrm{xcode}}(p), & p \in \{8,4,2\}, \\ 0, & p = 0, \end{cases} \tag{3}$$

where $e_{\mathrm{core}}$ is array-to-buffer energy per stored bit, $e_{\mathrm{down}}$ is downstream transfer energy per transmitted bit, and $E_{\mathrm{xcode}}(p)$ is the small conversion energy (zero for q8; FFN q2 includes comparison and mux costs). Since the dominant $p e_{\mathrm{down}}$ term scales with transmitted bits, lowering $p$ reduces bandwidth demand, transfer latency, and movement energy; KV-base and FFN-anchor costs are separate.

### *C. Block-Refreshed Query-Ranked KV Transport*

BRQ-KV keeps a canonical q8 residual master and adapts precision rather than sparsity (Figure 3).

**Canonical persistent representation.** After $N_{\min}$ finalized tokens, the *One-time pre-RoPE rank-two bases* stage computes fixed basis matrices $U^K, U^V$ ($r_{\mathrm{kv}} = 2$) for each layer and KV head from a streamed floating-point right Gram matrix, retaining only the top-two eigenvectors. Keys are factorized before rotary position embedding (RoPE), whereas Values are factorized directly. *Canonical low-rank-plus-residual encoding* defines coefficient $c_i = x_i U$, low-rank base $c_i U^\top$, and residual $r_i = x_i - c_i U^\top$ for $x_i \in \{k_i^{\mathrm{pre}}, v_i\}$. The *Persistent multi-view master* stores coefficient matrices $C^K, C^V$ (rows $c_i$), per-entry signed-INT8 residual codes, and fixed BF16 basis matrices $U^K, U^V \in \mathbb{R}^{d_h \times 2}$ ($d_h$ the per-head dimension) in DRAM. With $\mathcal{Q}_\tau$ denoting q$\tau$ residual decoding and $\mathcal{Q}_0(r) = 0$, the tier-$\tau$ view is

$$\hat{x}_i^{(\tau)} = c_i U^\top + \mathcal{Q}_\tau\left(r_i^{(8)}\right), \qquad \tau \in \{8,4,2,0\}. \quad (4)$$

q4/q2 remain read-time views and never overwrite q8; q0 drops only residual traffic and multiply-accumulate operations (MACs), so every prefix entry remains active through its low-rank base.

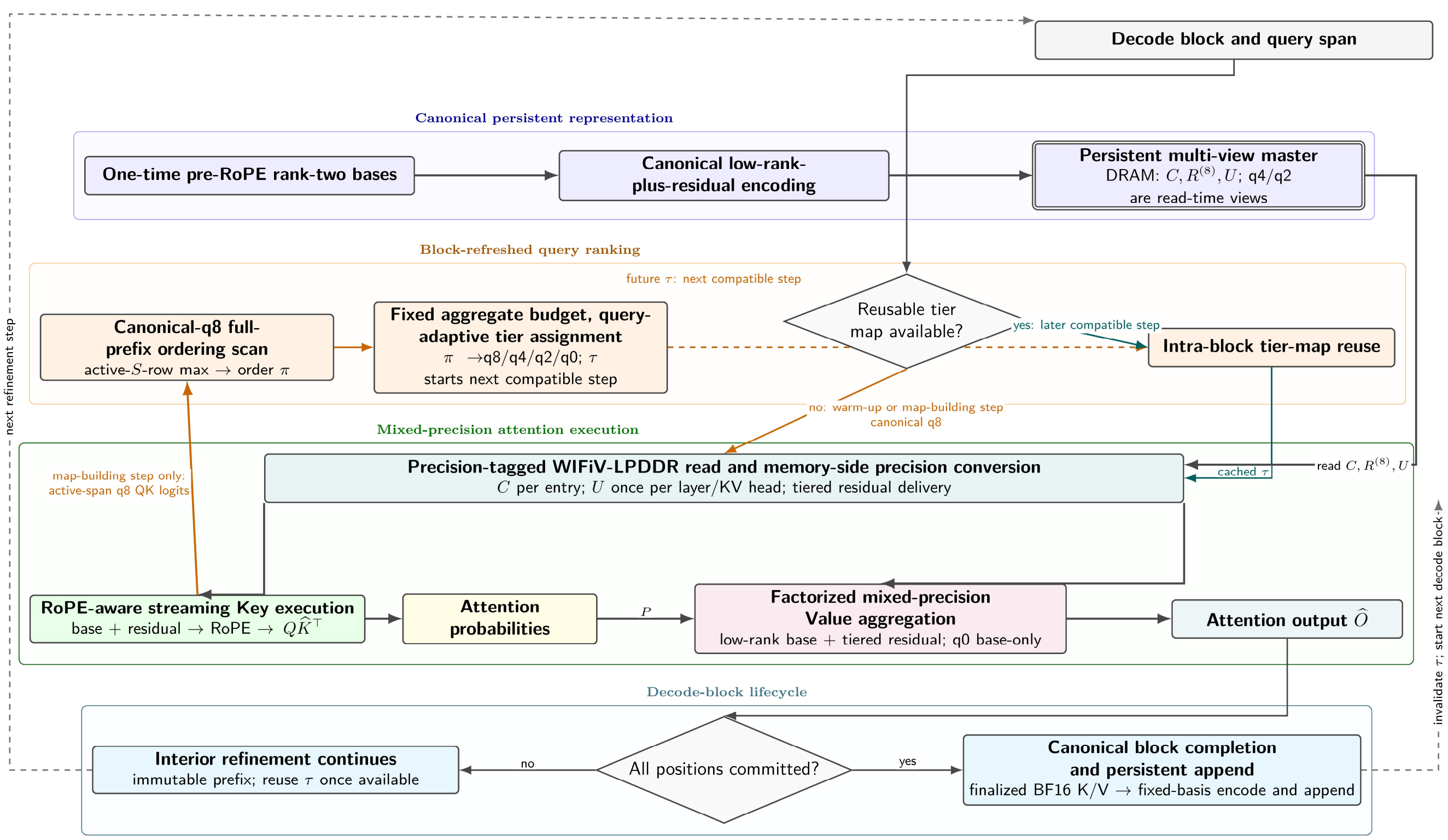

**Fig. 3.** BRQ-KV dataflow around the canonical q8 residual master: q8 warm-up, span-wise map construction, next-compatible-step reuse, and fixed-basis append. Orange and teal denote the q8 and reuse paths, respectively.

**Block-refreshed query ranking.** In Figure 3, *Decode block and query span* presents the current queries; *Reusable tier map available?* routes steps without a map through canonical q8, and later compatible steps through cached-map reuse. The first step of a new block is a ranking-free warm-up because its queries are least refined; the second step builds the first map. Each later span builds its map in its first step. A map-building step may evaluate all $B$ positions but ranks the prefix using only the current $S$-token query group. Its logits perform the *Canonical-q8 full-prefix ordering scan*; for layer $\ell$, KV head $h$, block $b$, query group $\Omega_{\ell h}^{b,s}$, prefix length $N_b$, and position-specific RoPE rotation $\mathcal{R}_i$,

$$\alpha_{\ell h i}^{b,s} = \max_{q \in \Omega_{\ell h}^{b,s}} \frac{\left\langle q, \mathcal{R}_i\left(c_i^K (U^K)^\top + \mathcal{Q}_8\left(r_i^{K,(8)}\right)\right)\right\rangle}{\sqrt{d_h}}, \quad (5)$$
$$\pi_{\ell h}^{b,s} = \operatorname{argsort}^{\downarrow}_{i \in [1, N_b]} \alpha_{\ell h i}^{b,s}.$$

*Fixed aggregate budget, query-adaptive tier assignment* partitions the prefix ordering $\pi$ into q8/q4/q2/q0 tiers using cumulative $\boldsymbol{\rho}$ boundaries. The reuse-step average residual width is $\bar{b}_{\mathrm{kv}} = 8\rho_8 + 4\rho_4 + 2\rho_2$ bits/entry, up to rounding. The resulting map $\tau$ starts at the next compatible step: the span remains q8 through map construction, then uses *Intra-block tier-map reuse*; block/span completion invalidates $\tau$.

**Mixed-precision attention execution.** As shown in *Precision-tagged WIFiV-LPDDR read and memory-side precision conversion*, each step fetches shared basis matrices $U$ once per layer/KV head into transient on-chip storage and streams per-entry $C$; warm-up/map-building steps request q8 residuals, while map hits follow $\tau$ (q8 direct, q4/q2 generated, q0 residual omitted). *RoPE-aware streaming Key execution* reconstructs each pre-RoPE Key to BF16, rotates it, and streams it without a full-dimensional Key cache into $Q\hat{K}^\top/\sqrt{d_h}$, where $Q$ stacks $\Omega_{\ell h}^{b,s}$. The *Attention probabilities* are $P = \mathrm{softmax}\left(Q\hat{K}^\top/\sqrt{d_h}\right)$. For $\mathcal{I}_p = \{i : \tau_i = p\}$ and decoded Value residuals $\hat{R}_{\mathcal{I}_p}^V = D_p Z_p$, *Factorized mixed-precision Value aggregation* computes

$$\hat{O} = (PC^V)(U^V)^\top + \sum_{p \in \{8,4,2\}} \left(P_{:,\mathcal{I}_p} D_p\right) Z_p, \quad (6)$$

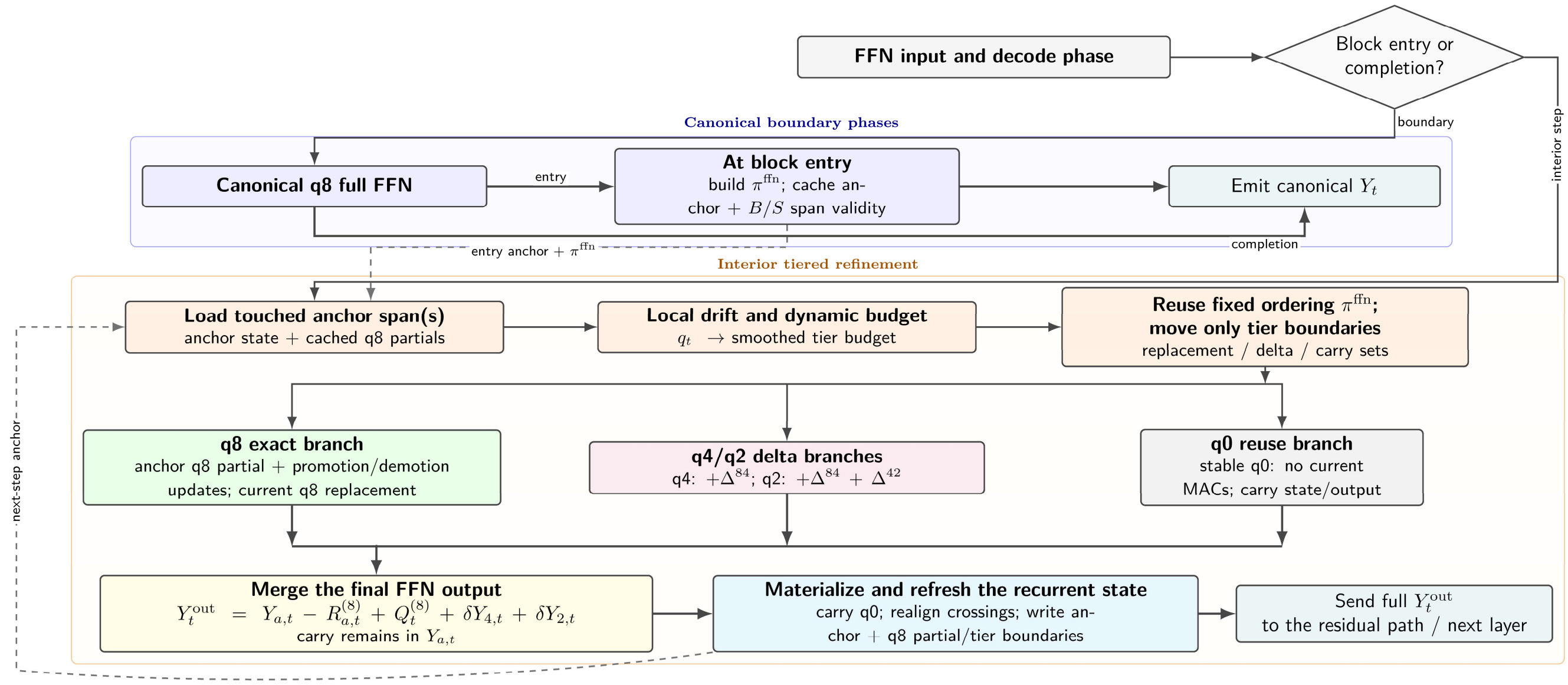

**Fig. 4.** DAT-FFN dataflow: canonical boundary execution, block-entry ordering, interior replacement/delta/carry, and per-step anchor refresh.

where $Z_8$ is direct; offline-fitted clip/shift and threshold maps form $Z_4$ and $Z_2$. Diagonal $D_p$ stores per-entry scales, with $D_4, D_2$ combining the q8 scale and fitted amplitude factor (Section 4.A).

In *Input-stationary array mapping*, the BF16 query tile is stationary on mixed-precision hardware adapted from prior work [10]. Keys are reconstructed to BF16 before RoPE, so compression saves memory/staging rather than QK MAC precision. Value residual codes enter at delivered q8/q4/q2 precision without reconstructing complete Value entries; the low-rank branch and probability tile $P$ are BF16, with $P$ stationary after softmax. Within each KV head, residual-tier buckets and the low-rank branch stream separately; q0 skips its residual, and floating-point partials are cast once. The *Attention output* $\hat{O}$ enters block refinement.

**Decode-block lifecycle.** While uncommitted positions remain, *Interior refinement continues* over an immutable prefix. *Canonical block completion and persistent append* reads that prefix through its low-rank base plus canonical-q8 residual, encodes finalized current-block K/V (still BF16) with the fixed basis matrices, and appends the rows; transient states are never committed.

## *D. Drift-Adaptive Tiered FFN*

**Core policy.** DAT-FFN fixes a block-entry channel order and maps per-step input drift to replacement, weight-view delta, or carry over cached gated-FFN state. Live activations stay BF16; approximation comes from selected views, carry, and one-step tier crossings, not a Taylor expansion. Figure 4 summarizes the phases.

**Offline canonical weight representation.** For $m \in \{g, u, d\}$ (Gate, Up, Down), standard GPTQ [11] supplies canonical q8 codes $Z_m^{(8)}$ and per-group/output scales $s_{g,o}^m$, with $\widehat{W}_{m,oi}^{(8)} = s_{g(i),o}^m Z_{m,oi}^{(8)}$, where $o$ indexes output channels and $g(i)$ denotes the GPTQ group containing input column $i$. Starting from this q8 master, DAT-FFN derives q4/q2 read-time views and adjacent corrections instead of storing low-bit weight copies. Let $\mathcal{M}_{m,4}$ and $\mathcal{M}_{m,2}$ denote the fixed q8-to-q4 and decoded-q4-to-ternary-q2 code maps. For $b \in \{4,2\}$, offline calibration selects each map's clamp and shift $k_{m,b}$, fits per-group/output decode factors $\alpha_{m,g,o}^{(b)}$, and fits rank-$r_{84}$ and rank-$r_{42}$ factors to the scale-weighted adjacent errors; all are fixed during inference:

$$\begin{aligned} \overline{Z}_{m,oi}^{(4)} &= \alpha_{m,g(i),o}^{(4)} 2^{k_{m,4}} \left[\mathcal{M}_{m,4}\left(Z_m^{(8)}\right)\right]_{oi}, \\ \overline{Z}_{m,oi}^{(2)} &= \alpha_{m,g(i),o}^{(2)} 2^{k_{m,2}} \left[\mathcal{M}_{m,2}\left(\overline{Z}_m^{(4)}\right)\right]_{oi}. \end{aligned} \tag{7a}$$

$$\begin{aligned} E_m^{84} &= Z_m^{(8)} - \overline{Z}_m^{(4)} \approx \Delta_m^{84} = C_m^{84}(V_m^{84})^\top, \\ E_m^{42} &= \overline{Z}_m^{(4)} - \overline{Z}_m^{(2)} \approx \Delta_m^{42} = C_m^{42}(V_m^{42})^\top. \end{aligned} \tag{7b}$$

$$\begin{aligned} \hat{Z}_m^{(8)} &= Z_m^{(8)}, \\ \hat{Z}_m^{(4)} &= \overline{Z}_m^{(4)} + \Delta_m^{84}, \\ \hat{Z}_m^{(2)} &= \overline{Z}_m^{(2)} + \Delta_m^{84} + \Delta_m^{42}. \end{aligned} \tag{7c}$$

$$\widehat{W}_{m,oi}^{(p)} = s_{g(i),o}^m \hat{Z}_{m,oi}^{(p)}, \qquad p \in \{8,4,2\}. \tag{7d}$$

The DAT-FFN residual ladder comprises decoded views (Eq. (7a)), adjacent q8→q4 and q4→q2 corrections (Eq. (7b)), their stacking (Eq. (7c)), and inherited GPTQ scaling (Eq. (7d)). The q2 view reuses the same $\Delta_m^{84}$ as q4 and adds only $\Delta_m^{42}$.

**Factorized runtime correction.** For projection $m$, let $\Delta Z = CV^\top$ denote either adjacent correction in Eq. (7b) and $H$ its input activation: $X$ for Gate/Up and the gated activation for Down. Since $\Delta Z$ is on the q8-code scale, $\Delta W_{oi} = s_{g(i),o}^m \Delta Z_{oi}$. For GPTQ group $g$ with input-column set $\mathcal{G}_g$, define $H_g = H_{:,\mathcal{G}_g}$, $V_g = V_{\mathcal{G}_g,:}$, $s_g = \left(s_{g,o}^m\right)_o$, and $C_g = \text{Diag}\left(s_g\right) C$. Then $H\Delta W^\top = \sum_g \left(H_g V_g\right) C_g^\top$ without materializing $\Delta W$. Main q$b$ code partials use $s^{m}_{g,o}\alpha^{(b)}_{m,g,o}2^{k_{m,b}}$; correction partials use only the GPTQ scale absorbed into $C_g$.

**Canonical boundary phases.** Each *FFN input and decode phase* routes block entry/completion through *Canonical q8 full FFN*: $G = X\widehat{W}_g^{(8)\top}$, $U = X\widehat{W}_u^{(8)\top}$, $S = SiLU(G)$, the gated FFN intermediate activation is $A = S \odot U$, and $Y = A\widehat{W}_d^{(8)\top}$; *Emit canonical* $Y_t$ returns $Y$. *At block entry*, cache $(X_a, G_a, U_a, S_a, Y_a)$ in $B$- or $S$-sized spans and order channels once by $\pi^{\text{ffn}} = \text{argsort}_j^{\downarrow}\left[\text{mean}_n|A_{a,nj}|\ \|\widehat{W}_d^{(8)}[:,j]\|_2\right]$; completion skips anchor construction.

**Interior tiered refinement.** An interior step begins with *Load touched anchor span(s)*, reading cached q8 output partials $Q^{(8)}_{a,t,s}$. With $\delta X_t = X_t - X_{a,t}$, *Local drift and dynamic budget* uses

$$q_t = \text{clip}_{[0,1]}\left(\frac{\|X_t - X_{a,t}\|_F}{\gamma_X \max\left(\|X_{a,t}\|_F, \epsilon\right)}\right), \quad (8)$$

and

$$\begin{aligned}\tilde{\rho}_{p,t} &= clip\left(\rho_p^{\min} + \beta_p q_t, \rho_p^{\min}, \rho_p^{\max}\right), \quad p \in \{8,4,2\},\\ \rho_{p,t} &= \eta\rho_{p,t-1} + (1-\eta)\tilde{\rho}_{p,t}, \qquad \rho_{0,t} = 1 - \textstyle\sum_{p\in\{8,4,2\}}\rho_{p,t},\end{aligned} \quad (9)$$

to *Reuse fixed ordering* $\pi^{ffn}$*; move only tier boundaries*. Here $\gamma_X$ normalizes drift, $\epsilon$ prevents division by zero, $\beta_p$ is the tier gain, and $\eta$ smooths the budget.

The *q8 exact branch* evaluates the current replacement set, giving $Q_t^{(8)}$. Define $\mathcal{H}(\mathcal{J}) = \left[S_{a,\mathcal{J}} \odot \left(X_a\widehat{W}_{u,\mathcal{J}}^{(8)\top}\right)\right]\widehat{W}_{d,\mathcal{J}}^{(8)\top}$ and promotion/demotion sets $\mathcal{J}^{\pm}$. The aligned anchor reference is

$$R^{(8)}_{a,t,s} = Q^{(8)}_{a,t,s} + \mathcal{H}\left(\mathcal{J}^{+}_{t,s}\right) - \mathcal{H}\left(\mathcal{J}^{-}_{t,s}\right). \quad (10)$$

$\mathcal{H}\left(\mathcal{J}^{\pm}_{t,s}\right)$ is evaluated only for intermediate channels entering or leaving q8 relative to the cached state of span $s$; thus the extra anchor-side work scales with $|\mathcal{J}^{+}_{t,s}| + |\mathcal{J}^{-}_{t,s}|$, not the full q8 set. For $p \in \{4,2\}$, the *q4/q2 delta branches* project $\delta X_t$ through Gate/Up with $\widehat{W}_g^{(p)}$ and $\widehat{W}_u^{(p)}$ (Eq. (7d)), use $\delta S = SiLU(G_a + \delta G) - S_a$, form $\delta A = (S_a + \delta S) \odot (U_a + \delta U) - S_a \odot U_a$, and apply Down with $\widehat{W}_d^{(p)}$ to obtain $\delta Y_{p,t}$. Same-tier updates are algebraically exact relative to the selected view; crossings use the previous tier anchor for one step and realign at refresh. The *q0 reuse branch* carries cached state and output contribution without current Gate/Up/Down weight access; only q0-boundary crossings realign at refresh.

In *Input-stationary array mapping*, the BF16 activation tile is stationary: $X_t$ or $\delta X_t$ for Gate/Up and the corresponding gated-activation tile ($A$ or $\delta A$) for Down. Gate/Up tier rows stream without reordering the activation tile; within each 128-column GPTQ group $\mathcal{G}_g$, a gather buffer compacts $\mathcal{G}_g \cap \mathcal{I}_p(t)$ and streams matching Down columns from $\widehat{W}_d^{(p)}$. Each main low-bit partial applies its fused scale once, unchanged q0 channels issue no current MACs, and correction branches accumulate separately.

**Merge and per-step anchor refresh.** Let $R^{(8)}_{a,t}$ and $Q^{(8)}_t$ concatenate touched-span partials; *Merge the final FFN output* gives

$$Y_t^{\text{out}} = Y_{a,t} - R^{(8)}_{a,t} + Q^{(8)}_t + \delta Y_{4,t} + \delta Y_{2,t}, \quad (11)$$

where carry remains in $Y_{a,t}$. In *Materialize and refresh the recurrent state*, nonzero tiers form the next $G/U/S/Y$ anchor, while unchanged q0 state and output contribution are carried; q0 refresh accesses canonical-q8 Down columns only at boundary crossings. The refresh pass commits $X_a$, $(G,U,S,Y)_a$, the next q8 partial, and current tier boundaries; $T = S$ patches one span and $T = B$ refreshes all. The *Send full* $Y_t^{out}$ *to the residual path / next layer* stage forwards the current-step output.

## IV. Evaluation

### *A. Evaluation Setup*

We use LLMET [12], a technology-calibrated cycle-level model of compute, on/off-chip memory, conversion, KV-basis/correction traffic, packing, metadata, and utilization. Fast-dLLM v2 1.5B/7B [3] map to Jetson Orin Nano 4GB/NX 16GB [6] with 7-nm arrays matched to platform throughput. LPDDR5 baselines provide 51.2/102.4 GB/s; WIFiV-LPDDR is conservatively modeled at 4 × (204.8/409.6 GB/s), not its architectural maximum. We report HumanEval, MBPP, GSM8K, MATH, and IFEval under multi-token generation [3]. Under the adopted Fast-dLLM v2 protocols, MMLU and GPQA instead use fixed-choice likelihood scoring and do not exercise the generation path [3].

All configurations share GPTQ-8 FFN and Q/K/V projection weights. *Baseline* uses LPDDR5 without BRQ-KV/DAT-FFN; *Compression* adds both algorithms and conversion logic; WIFiV-LPDDR changes only memory; *Full* combines both. Added logic and all compute, memory, conversion, packing, scale, metadata, and utilization costs are included. Each pair is normalized to Baseline; reported model factors are arithmetic means of the five Baseline/configuration ratios.

Decoding uses $B = 32$, $S = 8$, $r_{\text{kv}} = 2$, and $N_{\min} = 128$. A one-time offline controller pass on a fixed GSM8K-training calibration subset, with no test examples, produced the target $\bar{b}_{\text{kv}} = 4.6861$ residual bits/entry. Evaluation fixes this target and solves

$$\begin{aligned}\boldsymbol{\rho}_{\text{kv}}(d) &= (0.2 + 0.2d,\ 0.4 + 0.05d,\\ &\qquad 0.25 - 0.1d,\ 0.15 - 0.15d),\\ \bar{b}_{\text{kv}} &= 3.7 + 1.6d.\end{aligned}$$

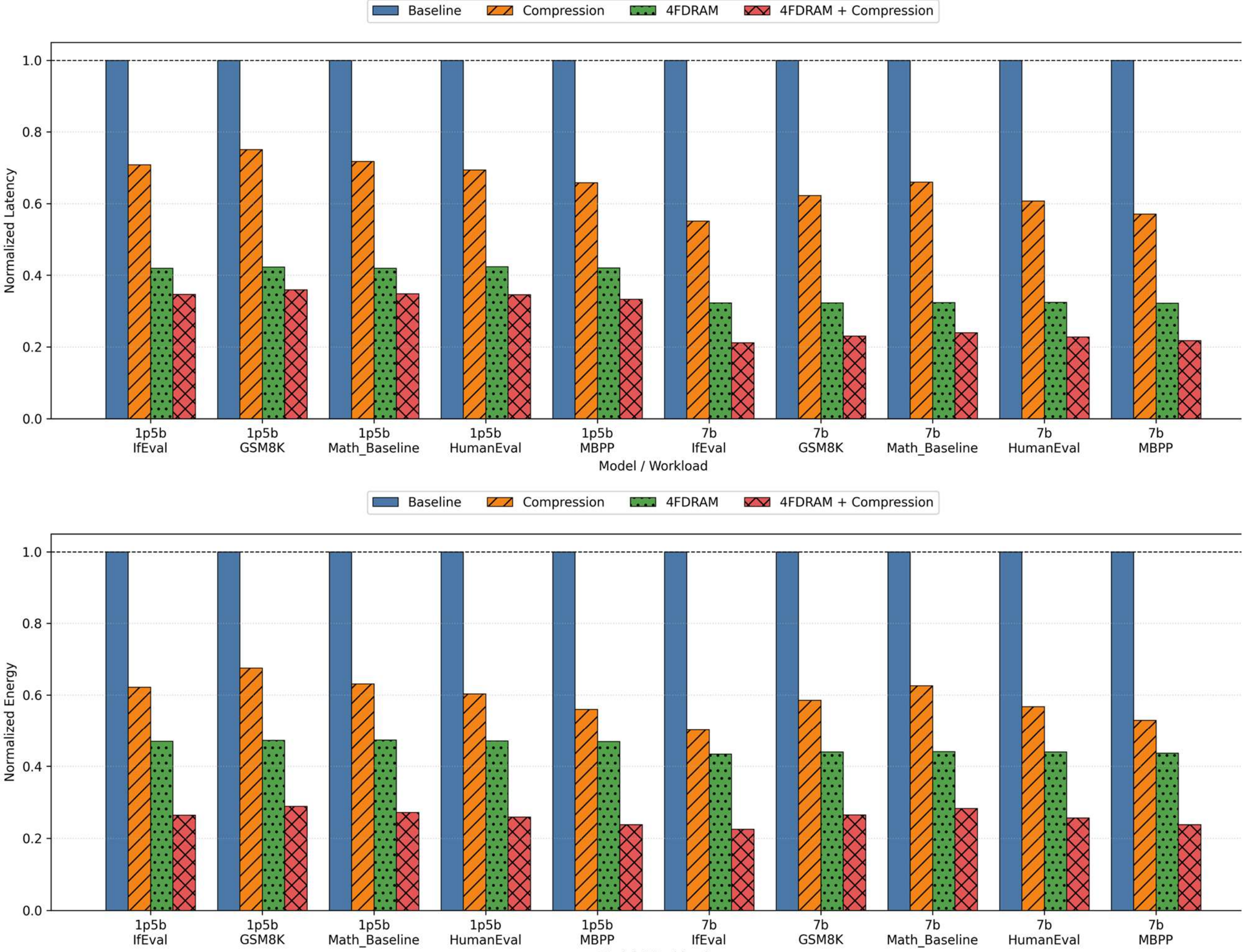

**Fig. 5.** Normalized end-to-end latency (top) and energy (bottom) for the four configurations of Section 4.A. 4FDRAM denotes WIFiV-LPDDR, 4FDRAM + Compression denotes Full, and Math_Baseline denotes MATH.

Solving gives $d \approx 0.6163$ and q8/q4/q2/base-only shares $(0.32326, 0.43082, 0.18837, 0.05755)$, fixed across models and benchmarks. The same subset fixes the KV clip/shift and threshold maps (read-time $Z_4/Z_2$) and the amplitude factors in $D_4/D_2$ (Eq. (6)).

DAT-FFN uses BF16 activations, $r_{84} = r_{42} = 8$, a 10% q8 bypass selected by scale-weighted q8-to-q4 reconstruction error, EMA factor $\eta = 0.7$, and drift $q_t$ with $\gamma_x = 1$, $\epsilon = 10^{-6}$. The q4/q2 maps and corrections are computed offline and fixed. Seven pre-EMA settings trade fidelity for traffic with KV fixed. Setting 0 uses q8/q4 only: $\tilde{\rho}_8^{(0)} = clip_{[0.05,0.07]}(0.05 + 0.20q_t)$, $\tilde{\rho}_4^{(0)} = 1 - \tilde{\rho}_8^{(0)}$, and $\tilde{\rho}_2^{(0)} = \tilde{\rho}_0^{(0)} = 0$. Settings 1-6 use $\tilde{\rho}_8 = 0.05 + 0.025q_t$, with q4 the remainder. For $j = 1,2,3$, Setting $j$ uses $\tilde{\rho}_0 = 0$ and $\tilde{\rho}_2 = c_j - 0.10q_t$, where $(c_1, c_2, c_3) = (0.15, 0.25, 0.35)$. Setting $j + 3$ uses $\tilde{\rho}_2 = 0.275$ and $\tilde{\rho}_0 = a_j - b_j q_t$, with $(a_j, b_j) = (0.10, 0.075)$, $(0.15, 0.10)$, and $(0.225, 0.125)$. Reported GSM8K/MATH/IFEval/HumanEval/MBPP settings are 0/3/2/3/4 for 1.5B and 4/2/6/4/5 for 7B.

### *B. Results*

All end-to-end measurements include prompt prefill; across the ten evaluated model-benchmark pairs, it accounts for at most 3% of either latency or energy. At the reported DAT-FFN settings, Figure 5 shows Full improves every pair. Arithmetic-mean energy-reduction/latency-speedup factors for 1.5B/7B are Compression $1.62\times/1.79\times$ and $1.42\times/1.67\times$; WIFiV-LPDDR $2.12\times/2.28\times$ and $2.37\times/3.09\times$; and Full $3.79\times/3.96\times$ and $2.88\times/4.44\times$. Full exceeds either component for every pair. Adding Compression atop WIFiV-LPDDR lowers energy more than latency because transferred-bit savings persist while compute and on-chip work limit latency.

## V. Conclusion

We co-designed WIFiV-LPDDR, BRQ-KV, and DAT-FFN for batch-one block-diffusion inference. For the evaluated 1.5B/7B models on modeled Jetson-class systems, Full provides arithmetic-mean energy-reduction factors of $3.79\times/3.96\times$ and arithmetic-mean latency speedups of $2.88\times/4.44\times$ at the reported DAT-FFN settings, while every corresponding compressed benchmark score drops by less than one absolute percentage point from its baseline.